\documentclass[a4paper]{article}

\usepackage[margin=2.5cm]{geometry}
\usepackage{graphicx}
\usepackage{subfigure}
\usepackage{changepage}
\usepackage{lipsum}
\usepackage{xcolor}
\usepackage[normalem]{ulem}

\newcommand{\del}[1]{\ifnum\value{diff}<1{\textcolor{red}{\sout{#1}}}\else{}\fi}
\newcommand{\add}[1]{\ifnum\value{diff}<1{\textcolor{blue}{#1}}\else{#1}\fi}

\begin{document}

\newcounter{diff}
\setcounter{diff}{0}

\title{The special point on the hybrid star mass--radius diagram and its multi--messenger implications}

\author{%
Mateusz Cierniak$^{1,a}$ and David Blaschke$^{a,b,c}$\\\mbox{}\\
{\small $^a$Institute of Theoretical Physics, University of Wroclaw, 50-204 Wroclaw, Poland}\\
{\small $^b$National Research Nuclear University (MEPhI), 115409 Moscow, Russia}\\
{\small $^c$Bogoliubov Laboratory of Theoretical Physics, Joint Institute for Nuclear Research, 141980 Dubna, Russia}
}
\footnotetext[1]{mateusz.cierniak@uwr.edu.pl}
\date{}

\maketitle

\begin{adjustwidth}{2 cm}{2 cm}

\textbf{Abstract.} We show the existence and investigate the location of the special point (SP) in which hybrid neutron star mass-radius (M-R) curves have to cross each other when they belong to a class of hybrid equation of state (EoS) constructed with generic constant--speed--of--sound (CSS) quark matter models for which the onset deconfinement is varied.
We demonstrate that for a three-parameter CSS model the position of the SP in the M-R diagram is largely independent of the choice of the hadronic EoS, but in dependence on the stiffness of the quark matter EoS it spans a region that we identify.
We find that the difference between the maximum mass and the SP mass depends on the mass at the onset of deconfinement so that an upper limit of $0.19~M_\odot$ for this difference is obtained from which a lower limit on the radius of hybrid stars is deduced.
Together with a lower limit on the radius of hadronic stars, derived from a class of reasonably soft hadronic EoS including hyperons, we identify a region in the M-R diagram which can be occupied only by hybrid stars.
Accordingly, we suggest that a NICER radius measurement on the massive pulsar PSR J0740+6620 in the range of 8.6-11.9 km would indicate that this pulsar is a hybrid neutron star with deconfined quark matter in the inner core.
\end{adjustwidth}

\section{Introduction}
\label{sec:1}

A large and comprehensive body of work exists on the equation of state of nuclear matter up to and exceeding the nuclear saturation density that has recently been reviewed in \cite{Oertel:2016bki}. 
Models, such as \cite{Akmal:1998cf,Gandolfi:2011xu,Typel:2009sy} are well fitted to existing data on nucleon-nucleon interactions, nuclear structure and nuclear matter saturation properties and can be readily used to derive the properties of neutron stars with a hadronic matter core. 
Such models are, however, blind to the quark substructure of strongly interacting matter, which in itself is a subject of many studies (cf. \cite{Bastian:2018wfl,Marczenko:2018jui,Marczenko:2019trv,Cierniak:2017dxr} and references therein).

The existing state--of--the--art field--theoretical description of strongly interacting matter is the theory of Quantum Chromodynamics (QCD). 
Ideally, one would want a description of strongly interacting matter in astrophysical systems derived directly from this theory. 
However, the first--principle calculations on the basis of the QCD Lagrangian have to exploit Monte-Carlo simulation techniques that up to now 
can be applied only in vacuum or at finite temperatures and densities corresponding to baryon chemical potentials not exceeding twice the temperature 
(cf. \cite{Fodor:2004nz,Aoki:2006we,Bazavov:2017dus,Gunther:2016vcp,Bazavov:2018mes}). 

These first--principle calculations predict a smooth crossover from hadronic matter to deconfined quark--gluon plasma (QGP) at a temperature of 
$156.5\pm 1.5$ MeV \cite{Bazavov:2018mes}. 
For applications to astrophysical systems, such as neutron stars, the temperature $T$ is well below the Fermi temperatures of nucleons and leptons, so that for the 
equation of state calculations can be performed assuming $T=0$.
With densities in excess of the nuclear saturation density, and due to the asymptotic freedom of QCD it is possible that a transition to QGP will also occur under such conditions, likely in the form of a first order phase transition.

The possibility of such a feature of neutron stars is widely discussed in the literature (cf. \cite{Haensel:2007yy} and reference therein). 
Specific properties of the QGP present in hybrid neutron stars and the phase transition itself could provide a supernova explosion mechanism triggered by the 
proto-neutron star formation arising from a strong first order phase transition within the collapsing star's core (cf. \cite{Sagert:2008ka,Yudin:2013rha,Fischer:2017lag}). 
Furthermore, the current efforts in providing so-called multi--messenger measurements of neutron star properties (cf. \cite{Antoniadis:2013pzd,Cromartie:2019kug,Abbott:2018exr,De:2018uhw,Miller:2019nzo,Bauswein:2020aag}) can potentially give us a glimpse into the QCD phase diagram at an area inaccessible by terrestrial heavy--ion collision experiments.

For that purpose, a theoretical description of strongly interacting matter with a 
phase transition between bound hadronic states and QGP is desperately needed. 
The state-of-the-art in this regard are classes of effective models, such as the tdBag \cite{Farhi:1984qu} and the CSS model 
\cite{Zdunik:2012dj,Alford:2013aca}, coupled to a separate hadronic equation of state via a Maxwell construction ensuring a first order phase transition. 
Such multi-phase models provide predictions on the M-R relations and the central density of stable neutron stars. 
But due to the presently large uncertainties in the simultaneous measurement of masses and radii, e.g., by the NICER experiment \cite{Miller:2019cac,Raaijmakers:2019qny} one cannot yet select a most favorable one among them (cf. \cite{Blaschke:2020qqj}). 
A recent study \cite{Yudin:2014mla} has found, that EoS from the class of CSS models share the property that all M-R curves, regardless of the hybrid star onset density, must cross a small region in the M-R diagram, the SP. 
We would like to extend this study to a wider range of EoS with additional microscopic features, such as vector repulsion, and investigate the possibility that the existence of a SP is a universal property of hybrid neutron star models. 
This could potentially provide a tool for the interpretation of current multi-messenger observations as signals for the existence of a hybrid neutron star branch.

The manuscript is organized as follows. 
In Section~\ref{sec:2} we present the EoS of the models used in this study. 
In Section~\ref{sec:3} we investigate the existence and the properties of the SP in each class of models. 
In Section~\ref{sec:4} we summarize our findings and present our conclusions.

\section{Equations of state}
\label{sec:2}

For the hadronic (confined quark) matter equations of state we choose from the class of density--dependent relativistic mean field models based on the 
"DD2" parametrization \cite{Typel:2009sy} with excluded volume effects according to \cite{Typel:2016srf}. 
They describe the properties of nuclear matter at low densities up to and slightly above nuclear saturation. 
For higher densities we have chosen three slightly different models for an approximate description of the thermodynamics of deconfined quark matter:
\begin{enumerate}
\item the CSS model  \cite{Alford:2013aca} which is similar to the one used in \cite{Yudin:2013rha},
\item a bag model inspired by \cite{Farhi:1984qu}, 
but modified by a pressure and energy--density shift corresponding to the formation of a vector condensate related to repulsive vector--channel interaction on the quark level
\item the novel vBag model \cite{Klahn:2015mfa,Cierniak:2018aet,Klahn:2017exz,Cierniak:2019hhe,Salinas:2019fmu}, which combines the previous bag model with a non--trivial correlation between the hadronic and quark phases in order to impose a simultaneous onset of chiral symmetry restoration and deconfinement.
\end{enumerate}

The equation of state of the CSS model \cite{Alford:2013aca} postulates a relation between thermodynamic pressure $p$ and the energy density  $\epsilon$ 
of the form
\begin{equation}
    \epsilon = \epsilon_0+\frac{p}{c^2_s},
\end{equation}
where $c^2_s$ is the speed of sound squared and $\epsilon_0$ is a constant energy density shift. 
The pressure, described as a function of the baryon chemical potential $\mu_B$, is
\begin{equation} \label{eq2}
    p(\mu_B)=A\mu^{1+\beta}_B-B,
\end{equation}
which can be inverted to give
\begin{equation}
    \mu_B(p)=\left(\frac{p+B}{A}\right)^{1/(1+\beta)},
\end{equation}
where $A$, $B$ and $\beta$ are constant model parameters. The derivative of the pressure with respect to the baryon chemical potential is
\begin{equation}
    \frac{\partial p(\mu_B)}{\partial\mu_B}=n_B(\mu_B)=(1+\beta)A\mu^\beta_B.
\end{equation}
Using the relation $\epsilon=\mu_Bn_B-p$ we obtain the energy density as a function of baryon chemical potential,
\begin{equation}
\epsilon = B + \beta A\mu^{1+\beta}_B.
\end{equation}
The relation between energy--density and pressure takes the form
\begin{equation}
\epsilon=\beta p + (1+\beta)B,
\end{equation}
which shows us, that $c^2_s=1/\beta$ and $\epsilon_0=(1+\beta)B$. 
The parameter A can be varied without changing the relation of pressure and energy density, but it should be chosen such, that the jump in baryon density at the phase transition is not negative.

Both of the bag models \cite{Farhi:1984qu,Cierniak:2018aet} used here start from the thermodynamics of a gas of non--interacting fermions \cite{Kapusta:2006pm} 
with the dispersion relation $E(p)=\sqrt{{p}^2+m^2}$,
\begin{eqnarray}
    p_{fg}(\mu,T)&=&\frac{N_cN_f}{\pi^2}\int d{p}\frac{{p}^4}{3E(p)}\left(\frac{1}{1+e^{(E(p)-\mu)/T}}+\frac{1}{1+e^{(E(p)+\mu)/T}}\right),\\
    n_{fg}(\mu,T)&=&\frac{N_cN_f}{\pi^2}\int d{p}{p}^2\left(\frac{1}{1+e^{(E(p)-\mu)/T}}-\frac{1}{1+e^{(E(p)+\mu)/T}}\right),\\
    \epsilon_{fg}(\mu,T)&=&\frac{N_cN_f}{\pi^2}\int d{p}{p}^2E(p)\left(\frac{1}{1+e^{(E(p)-\mu)/T}}+\frac{1}{1+e^{(E(p)+\mu)/T}}\right),\\
\end{eqnarray}
and add a constant pressure shift, the bag constant $B$,
\begin{eqnarray}
    p(\mu)&=&p_{fg}(\mu,0)-B,\\
    n(\mu)&=&n_{fg}(\mu,0),\\
    \epsilon(\mu)&=&\epsilon_{fg}(\mu,0)+B.
\end{eqnarray}
A well known feature of such models is their inability to support a high mass neutron star consistent with observations (\cite{Antoniadis:2013pzd,Cromartie:2019kug}). 
In order to remedy the problem, repulsive vector interactions are postulated to modify the quark equation of state beyond the free gas approximation. 
In this study, we follow a construction based on \cite{Klahn:2015mfa}, which ensures that the speed of sound does not exceed the causality limit $c^2_s=1$,
\begin{eqnarray}
\label{bag1}
    p(\mu)&=&p_{fg}(\mu^*,0)+\frac{K_v}{2}n^2(\mu)-B,\\
    n(\mu)&=&n_{fg}(\mu^*,0),\\
    \epsilon(\mu)&=&\epsilon_{fg}(\mu^*,0)+\frac{K_v}{2}n^2(\mu)+B,\\
    \mu&=&\mu^*+K_vn_{fg}(\mu^*,0).
\end{eqnarray}
The additional model parameter, $K_v$ is related to the vector quark current interaction and can be related to the gluon mass scale (cf. \cite{Cierniak:2018aet}). 
The above equation of state fully defines the model, which later will be denoted as the ordinary bag model.

The model vBag, which also belongs to the class of thermodynamic bag models, attempts to phenomenologically account for both quark confinement and high density chiral symmetry restoration. It does so by redefining the bag constant of Eq. (\ref{bag1}) as the flavor-dependent thermodynamic shift associated with the vacuum pressure contribution of the chiral condensate. 
It can therefore be related to the dressed quark mass in vacuum, $B^f_\chi=P(M_f,\mu_f=0)-P(m_f,\mu_f=0)$ \cite{Cahill}, where $M_f$ is the full dressed quark mass and $m_f$ is the current quark mass (cf. \cite{Cierniak:2017dxr} and reference therein).
\begin{eqnarray}
    p_f(\mu_f)&=&p_{f,fg}(\mu^*_f,0)+\frac{K_v}{2}n^2_f(\mu_f)-B_{f,\chi},\\
    n_f(\mu_f)&=&n_{f,fg}(\mu^*_f,0),\\
    \epsilon_f(\mu_f)&=&\epsilon_{f,fg}(\mu^*_f,0)+\frac{K_v}{2}n^2_f(\mu_f)+B_{f,\chi},\\
    \mu_f&=&\mu^*_f+K_vn_{f,fg}(\mu^*_f,0).
\end{eqnarray}
The model additionally assumes a simultaneous onset of chiral symmetry restoration and deconfinement, which is imposed by introducing an additional pressure shift $B_{dc}$ equal to the pressure of the hadronic phase at the critical chiral chemical potential (i.e. $\mu_{f,\chi}$ such, that $\sum_f p_f(\mu_{f,\chi})=0$). 
This by construction ensures, that the phase transition occurs at $\mu_{f,\chi}$ and the full thermodynamic description of chiral physics is consistently described across both phases,
\begin{eqnarray}
    p_Q(\mu_B)&=&\sum_f p_f(\mu_f)+B_{dc},\\
    n_Q(\mu_B)&=&\sum_f n_f(\mu_f),\\
    \epsilon_Q(\mu_B)&=&\sum_f \epsilon_f(\mu_f)-B_{dc}.
\end{eqnarray}
The specifics of this alternative bag model lies in its non--trivial connection to the hadronic EoS in the form of the $B_{dc}$ parameter. 
In all previous two--phase models, the phases were independent, apart from the Maxwell transition interface. 
The vBag model offers an alternative, with an explicit hadronic impact of the quark equation of state. 
Since the aim of this study is to investigate a shared property of the hybrid neutron star branches derived using a two-phase model, 
the impact of such a non-trivial connection warrants investigation.

\section{The hybrid star EoS special point}
\label{sec:3}

\begin{figure}
  \subfigure[]{
  \resizebox{0.5\columnwidth}{!}{%
  \includegraphics[width=\linewidth]{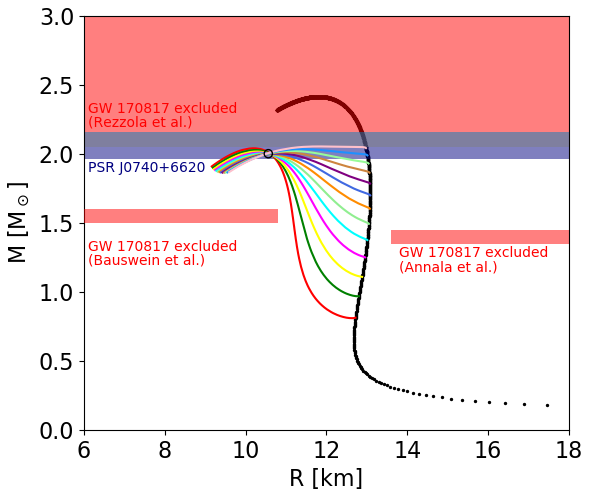}
  }
  \resizebox{0.5\columnwidth}{!}{%
  \includegraphics[width=\linewidth]{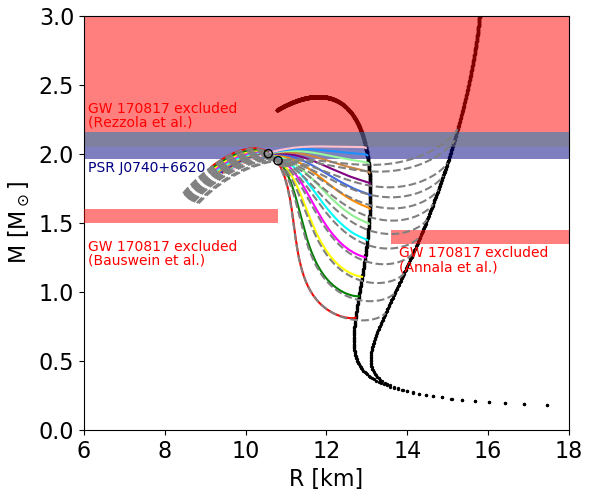}
  }
  }
  \subfigure[]{
  \resizebox{0.5\columnwidth}{!}{%
  \includegraphics[width=\linewidth]{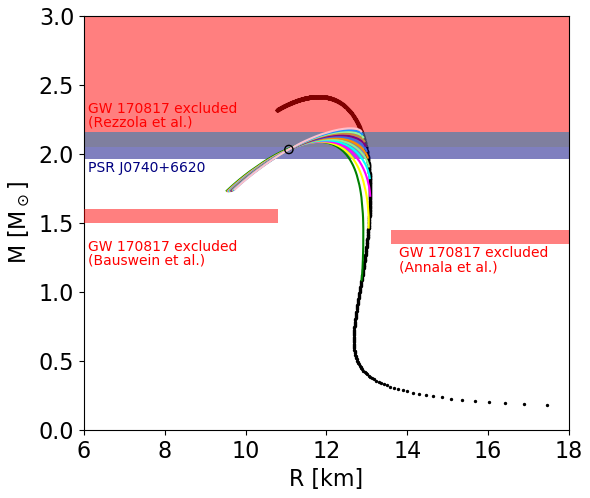}
  }
  \resizebox{0.5\columnwidth}{!}{%
  \includegraphics[width=\linewidth]{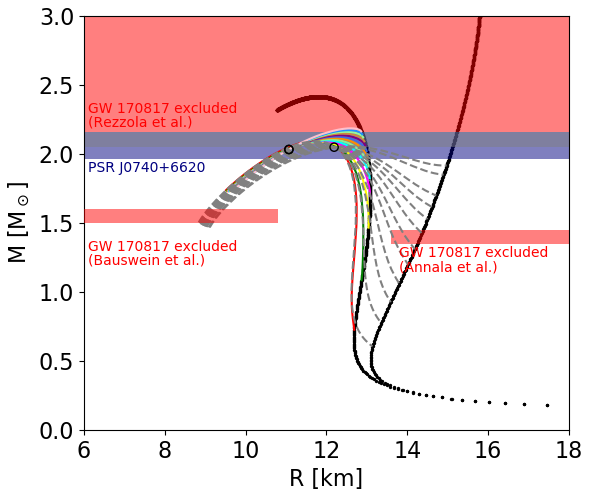}
  }
  }
  \subfigure[]{
  \resizebox{0.5\columnwidth}{!}{%
  \includegraphics[width=\linewidth]{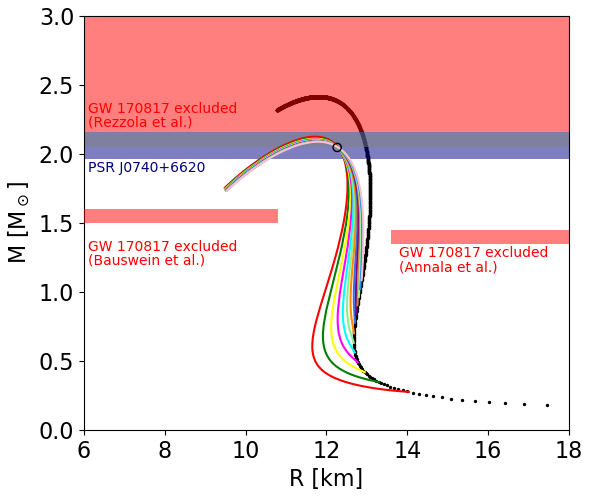}
  }
  \resizebox{0.5\columnwidth}{!}{%
  \includegraphics[width=\linewidth]{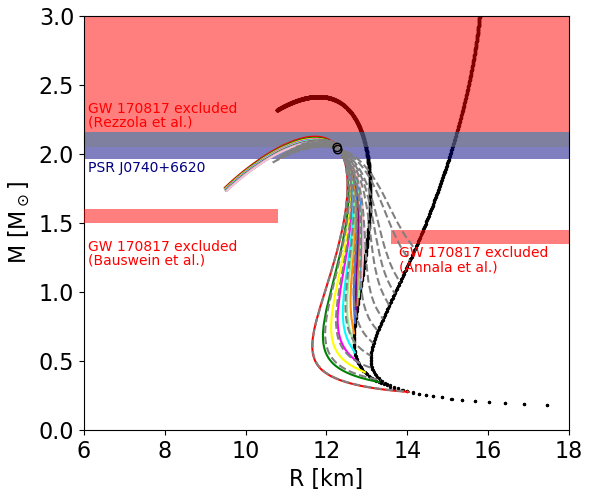}
  }
  }
  \caption{Mass--radius relations of neutron stars for different EoS, (a) - the CSS model, (b) - ordinary bag model, (c) - vBag, coupled to the hadronic EoS DD2p0 (left) and both DD2p0 and DD2p80 (right). Naming convention consistent with \cite{Typel:2016srf}. The special points are marked by black circles. The gray and blue bands correspond to $68.3\%$ and $95.4\%$ credibility intervals of the Shapiro--delay measurement of pulsar's PSRJ0740+6620 mass \cite{Cromartie:2019kug}. Red bands are regions excluded by the analysis of the signal from the neutron star merger event GW170817 according to Bauswein et al. \cite{Bauswein:2017vtn}, Annala et al. \cite{Annala:2017llu} and Rezzolla et al. \cite{Rezzolla:2017aly}.}
  \label{fig1}
\end{figure}

The existence of a SP in the M-R diagram of hybrid neutron stars, observed in \cite{Yudin:2014mla} is a feature visible in many other studies 
(for example, Fig.~2 of \cite{Blaschke:2010vd}, Fig.~14 of \cite{Kaltenborn:2017hus} or Fig.~2 of \cite{Blaschke:2020qqj}), including models using non--Maxwell phase transitions, see Fig.~8 of \cite{Blaschke:2020qqj}.
Indeed, for all the models described in the previous section, such a point does exist (examples can be seen in Fig.\ref{fig1}).
We need, however, to make a short remark on the use of the term "point" in reference to this feature. 
As illustrated in the original study (Fig. 5 of \cite{Yudin:2014mla}), there is no strict intersection between all hybrid M-R curves, but rather a very narrow region to which all of them converge. For that reason, the special point masses and radii, given in this study, should be considered approximate, along with any implications related to the multi-messenger constraints. Despite this shortcoming, we would argue that this feature does provide a useful mechanism for evaluating the agreement of a chosen hybrid star model with observations.

An example of robustness of the special point (SP) can be seen in the right panels of Fig.~\ref{fig1}, where the position of the SP is compared for different hadronic EoS coupled to the respective quark matter models. Of particular note are the panels depicting the difference in the vBag model (Fig.~\ref{fig1}c), which due to its non-trivial connection between quark and hadron EoS, exemplifies the worst-case scenario of the impact of the hadronic EoS on the SP position. 
We find, even in this case, that the change in mass and radius is less than $10\%$.

The SP doesn't always belong to the stable part of the M-R relation (as evident from Fig.~\ref{fig1}). It can be on the unstable branch when the QM phase transition occurs at rather high masses and with low latent heat. We would conjecture, that, for the more realistic EoS with a phase transition, the SP must appear on the stable part of the NS M-R plot. Studies, such as \cite{Blaschke:2020qrs}, suggest that nuclear matter stiffens at high density as a result of the hadron substructure effects. This naturally favors a phase transition with a relatively large latent heat and would result in a hybrid EoS with the SP on the stable hybrid star branch. Examples of this are evident in panels (a) and (c) of Fig.~\ref{fig1} and in the hybrid star models using advanced quark matter approaches in the literature, e.g. in \cite{Blaschke:2020qqj,Blaschke:2010vd,Kaltenborn:2017hus}.

\begin{figure}[!t]
\hspace{4.3em}
  \resizebox{0.7\columnwidth}{!}{%
  \includegraphics[width=\linewidth]{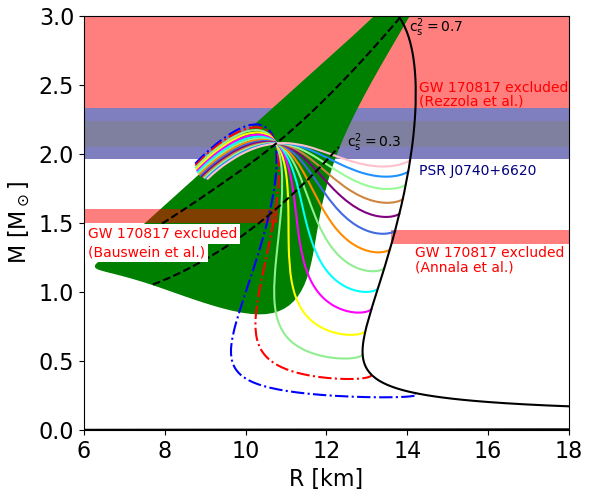}
  }
  \caption{The possible positions of the SP on the mass--radius plot (green area) for the CSS model with $c^2_s=0.7$. The colored lines represent an example parametrisations and the resulting SP point with dash--dotted lines showing parametrizations which violate the constraint from \cite{Bauswein:2017vtn}. Dashed black lines show the positions of SP's for $c^2_s=0.7$ and $c^2_s=0.3$.}
  \label{fig2}
\end{figure}

We would now like to focus on the CSS model, which was used in the original study by Yudin et al. \cite{Yudin:2014mla}. 
It is clearly established, that the SP is insensitive to a change in the quark matter onset density. 
The same is not true for a change of the speed of sound or the pressure slope (the A parameter of Eq.~(\ref{eq2})). 
Both of these quantities have a large impact on the shape of the hybrid star M-R relation. 
Fig.~\ref{fig2} illustrates the possible positions of the SP resulting from the change of these two parameters.
The SP can take values well above the 2 $M_\odot$ constraint of \cite{Cromartie:2019kug}, and the maximum mass of a hybrid star can be in excess of this value. In fact, as evident from this figure, for a hybrid branch with the so--called twin feature (i.e. with stable hybrid stars of lower masses than the heaviest purely hadronic neutron stars) it is a requirement.
\begin{figure}[!h]
\hspace{4.3em}
  \resizebox{0.7\columnwidth}{!}{%
  \includegraphics[width=\linewidth]{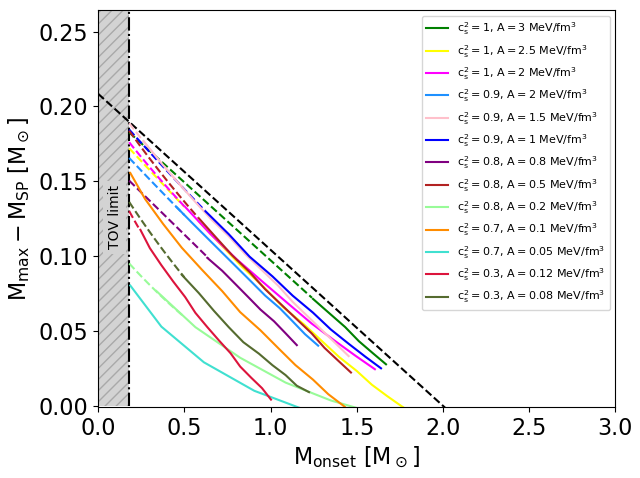}
  }
  \caption{The relation between the onset mass of hybrid neutron stars and the difference of the maximum neutron star mass and the SP mass for different parametrizations of the CSS model. The dashed line represents the upper limit on the interpolated maximum mass gain above the SP mass.}
  \label{fig3}
\end{figure}

\begin{figure}[h]
\hspace{4.3em}
  \resizebox{0.7\columnwidth}{!}{%
  \includegraphics[width=\linewidth]{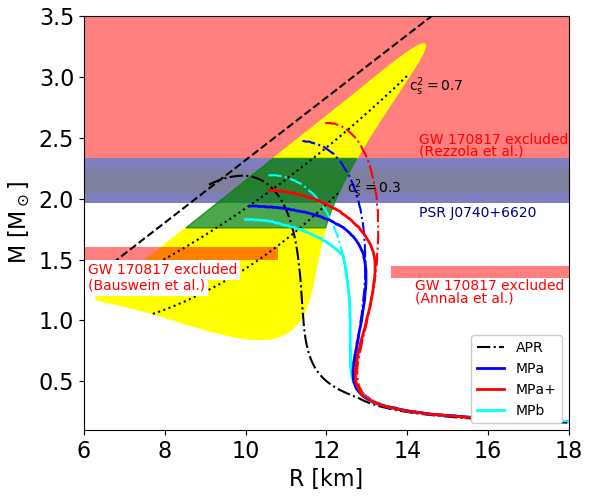}
  }
  \caption{The SP range allowed (green), and excluded (yellow) by the 2 $M_\odot$ constraint. 
The black dashed line shows the upper limit of masses that can be reached for a given radius of the hybrid star configurations.
The red, blue and cyan lines show the M-R relation for the hadronic EoS that are obtained with realistic baryon-baryon two-body and three-body interaction augmented by a repulsive short-range multi-pomeron (MP) exchange potential according to \cite{Yamamoto:2017wre,Yamamoto:2015lwa}. The solid lines represents EoS parametrizations with hyperons, while dash-dotted lines represent EoS with only non-strange baryons.} The black line is obtained for the APR EoS \cite{Akmal:1998cf}.
  \label{fig4}
\end{figure}

The SP, as a phenomenological tool, has the most utility in discussing the twin phenomenon. 
Since it must correspond to a stable hybrid star, it must obey all multi--messenger observational constraints. 
This means the SP area of Fig. \ref{fig2} must be limited to areas not excluded by the signals from GW170817. 
Additionally, the hybrid branch must reach the 2 $M_\odot$ band. 
The relation between the maximal mass gain above the SP mass ($M_{\rm max} - M_{\rm SP}$) and the hybrid star onset mass $M_{\rm onset}$ 
for deconfinement is shown in Fig.~\ref{fig3} for different parametrizations of the CSS model.
From Fig.~\ref{fig3} we can deduce a fit formula for the upper limit (dashed line) of the maximum mass as
\begin{equation}
\label{M-limit}
M_{\rm max} = M_{\rm SP} + 0.208~M_\odot - 0.104~M_{\rm onset}.
\end{equation}
Taken together with the present observational lower limit on the maximum mass of pulsars (the lower limit of the $95.4~\%$ credibility interval) of  
$2.14-0.18$ M$_\odot$ for PSR J0740+6620 \cite{Cromartie:2019kug},
this relation tells us that the maximum mass cannot be larger than $M_{\rm SP}+0.19$ or alternatively, that any parametrization of the EoS resulting in 
$M_{SP}$ below 1.77 $M_\odot$ will be in disagreement with observations. 
This is only true for a very early onset of deconfinement at the lower limit of the mass of neutron stars. 
With additional constraints on the quark matter onset density, this mass constraint becomes more strict. 
An example for the modified SP region is illustrated in Fig. \ref{fig4} by the green hatched region.
The relation (\ref{M-limit}) allows to predict a limiting M-R relation for hybrid stars which is shown as the dashed line in Fig.~\ref{fig4}.

As a possible implication the SP region has on signals from observations, one can suggest the region between the lower limit on neutron star (viz. hybrid star) radii and the lower limit on radii of hadronic neutron stars should be populated by hybrid neutron stars with deconfined quark matter in their cores. 
Should the APR EoS be considered as the softest realistic hadronic neutron star? We suggest that the predictions of the APR EoS for the behavior of nuclear matter at supranuclear densities should not be trusted, and therefore the corridor for identifying hybrid stars in the M-R diagram becomes wider, see the curves labelled with "MP" in Fig.~\ref{fig4}.
This argument is based on a work by Yamamoto et al. \cite{Yamamoto:2017wre} who have shown that realistic nucleon-nucleon (NN) potentials like the 
Argonne V18 (AV18) or the extended soft core (ESC) interactions fail to reproduce the large-angle differential cross section for $^{16}$O -- $^{16}$O scattering at $E_{Lab}=70$ AMeV. But with an additional repulsive, density-dependent multi-pomeron exchange interaction, the data is described very well. 
This addition to the NN interaction results in very good saturation properties and a stiffening of the nuclear EoS at supernuclear densities which allows to solve the hyperon puzzle \cite{Yamamoto:2015lwa} and leads to a shift of the M-R curves of the corresponding solutions of the TOV equations to larger radii, see the curves labelled MPa, MPa+ and MPb in Fig.~\ref{fig4}.
We choose the curve corresponding to the MPa+ parametrization, as it is the only multi-pomeron EoS in agreement with the PSR J0740+6620 mass measurement despite a high density onset of hyperons (thin red line on Fig.\ref{fig4}), and consider it as the lower limit of hadronic NS radii. All other potential-model based hypernuclear matter EoS suffer the hyperon puzzle problem and cannot be considered. We therefore postulate, that a NICER measurement of the PSR J0740+6620 radius between 8.6-11.9 km would be a strong indication of a quark matter core.

In deriving this range we are deliberately ignoring a low radius constraint proposed in \cite{Koppel:2019pys}.
The reason for this is the use of exclusively hadronic EoS and deriving the radius limit from an observed relation between the threshold mass of a prompt collapse of binary NS collision and the maximum NS compactness. Both of these quantities are sensitive to the features of the EoS and can significantly change due to a phase transition to deconfined quark matter (cf. \cite{Bauswein:2020aag}). For that reason we choose a more conservative limit proposed in \cite{Bauswein:2017vtn}. It is not free from this flaw but represents a systematic analysis of a wide range of different hadronic EoS, therefore we assume it holds in general. As evident in Fig.~\ref{fig2}, it affects the hybrid branch by imposing a lower limit on the quark matter onset. A systematic analysis of this effect in relation to the SP is beyond the scope of this paper, and merits further study.

\section{Conclusions}
\label{sec:4}

We have shown the existence and investigated the properties the special point, a characteristic feature shared by a wide range of hybrid neutron star equations of state from the class of two-phase models that allow a systematic variation of a quark matter parameter determining the onset of deconfinement. 
We have demonstrated that the position of this special point is only marginally depending on the underlying hadronic equation of state.
We have drawn conclusions on the allowed region for the location of the special point in the M-R diagram of a hybrid neutron star, derived using the CSS model, which would be in agreement with multi-messenger observational constraints. 
We define a corridor in the M-R plane that can be used to determine if subsequent mass and radius measurements characterize the target as a hybrid neutron star.
As an example, we suggest that PSR J0740+6620 with a $2\sigma$ lower limit on its mass at $1.96~M_\odot$ would be a candidate for a hybrid star with quark matter core when the measurement of its radius by, e.g., 
the NICER experiment would fall  in the range of 8.6-11.9 km.
Such a conclusion crucially depends on our knowledge of the lower limit for the radius of purely hadronic neutron stars models in the M-R diagram, but its importance merits further studies of the special point as a general feature of hybrid neutron star models.

\section*{Acknowledgements}

We thank Cole Miller for his remarks regarding the draft of this paper and Andreas Bauswein for his insightful comments on the minimal radius constraint from the threshold of black-hole formation. 
D.B. acknowledges discussions with Tom Rijken, Hajime Togashi and Yasuo Yamamoto on their work in Ref. \cite{Yamamoto:2017wre} and \cite{Yamamoto:2015lwa}. 
D.B. received support from the Polish National Science Centre under Grant Number UMO-2019/33/B/ST9/03059 and from the MEPhI Academic Excellence Program under contract number 02.a03.21.0005.

\end{document}